*Article*

# Estimation of Grüneisen parameter of high-entropy-alloy-type functional materials

**Fysol Ibna Abbas[1,2], Yuki Nakahira[1], Aichi Yamashita[1], Md. Riad Kasem[1], Miku Yoshida[1], Yosuke Goto[1], Akira Miura[3], Kensei Terashima[4], Ryo Matsumoto[4,5], Yoshihiko Takano[4], Chikako Moriyoshi[6], and Yoshikazu Mizuguchi[1*]**

[1] Department of Physics, Tokyo Metropolitan University, 1-1, Minami-osawa, Hachioji 192-0397, Japan
[2] Department of Electrical & Electronic Engineering, City University, Khagan, Birulia, Savar, Dhaka-1216, Bangladesh
[3] Faculty of Engineering, Hokkaido University, Kita 13 Nishi 8, Sapporo 060-8628, Japan
[4] International Center for Materials Nanoarchitectonics (MANA), National Institute for Materials Science, Tsukuba, Ibaraki 305-0047, Japan
[5] International Center for Young Scientists (ICYS), National Institute for Materials Science, Tsukuba, Ibaraki 305-0047, Japan
[6] Graduate School of Advanced Science and Engineering, Hiroshima University, 1-3-1 Kagamiyama, Higashihiroshima 739-8526, Japan
* Correspondence: mizugu@tmu.ac.jp

**Abstract:** In functional materials like thermoelectric materials and superconductors, the interplay between functionality, electronic structure, and phonon characteristics is one of the key factors to improve functionality and to understand the mechanisms. In the first part of this article, we briefly review investigations on lattice anharmonicity in functional materials by Grüneisen parameter ($\gamma_G$). One can find that the $\gamma_G$ can be a good scale for large lattice anharmonicity and for detecting a change in anharmonicity amplitude in functional materials. Then, we show original results on estimation of $\gamma_G$ for recently-developed high-entropy-alloy-type (HEA-type) functional materials with a layered structure and a NaCl-type structure. As a common trend between those two systems with two- and three-dimensional structures, we find that $\gamma_G$ increases by a slight increase in configurational entropy of mixing ($\Delta S_{mix}$), and then $\gamma_G$ decreases with increasing $\Delta S_{mix}$ in high-entropy region.

**Keywords:** Grüneisen parameter; lattice anharmonicity; high-entropy alloy; superconductors; thermoelectric materials

## 1. Introduction

### 1.1. Thermoelectric materials and superconductors

To solve energy problem and to prevent climate change, the development of effective energy-creating and energy-saving technologies are carrying significant tools in the last two decades. One of the solutions to those issues is thermoelectric (TE) materials and modules [1,2]. TE modules allow direct conversion of unused thermal energy into useful electrical power, which can help to reduce carbon dioxide emissions and contribute to a more sustainable society. However, the performance of TE devices needs to be further improved for practical application, and thus, development of new TE material that can allow us to fabricate high performance TE devices is needed. To estimate the performance of TE materials, dimensionless figure-of-merit ($ZT$), which is calculated by the following formula (1), is essential.

$$ZT = \frac{S^2\sigma T}{\kappa} = \frac{S^2\sigma T}{\kappa_{el}+\kappa_{ph}} \quad (1)$$

where $S$, $\sigma$, $\kappa$, and $T$ are Seebeck coefficient, electrical conductivity, thermal conductivity, and absolute temperature, respectively. Generally, $\kappa$ is considered as the total of $\kappa_{el}$ and $\kappa_{ph}$, which are contributions from electron (mobile carrier) and phonon in the total $\kappa$. Since $\kappa_{el}$ can be controlled by the modification of the electronic transport properties, as described by the Wiedemann-Franz law [3], finding a material with essentially low $\kappa_{ph}$ with a larger $S^2\sigma$ has been one of the strategies for developing high-$ZT$ materials, typically with $ZT > 1$ [1]. One of the key strategies for improving TE properties is creating a layered structure. Several layered compounds, such as $Bi_2Te_3$, Co oxides, and $CsBi_4Te_6$, have exhibited high thermoelectric performance [4–7]. In layered systems, low-dimensional electronic states, a structure composed of stacking of sheets, and/or a large unit cell could be advantageous for producing high TE properties. In addition, nano-structuring and band convergence have been found to be effective to improve $ZT$ in TE materials such as PbTe [2,8]. Another way to reduce $\kappa_{ph}$ is the use of lattice anharmonicity [9–12]. The strategy is relatively new, but various TE materials have been discovered to exhibit a high TE performance [13–15]. Recently, the introduction of the effect of high-entropy states in TE materials has been developed [2,16–20]. Due to the introduced disorder, $\kappa_{ph}$ is expected to be largely suppressed, and high $ZT \sim 2$ is observed in high-entropy-alloy-type (HEA-type) chalcogenides [19]. However, the essential strategy to obtain a low $\kappa_{ph}$ by entropy control or the affection to anharmonicity have not been established. Therefore, in this article, we addressed this issue on HEA-type TE materials using Grüneisen parameter ($\gamma_G$), which will be reviewed in the next section.

Another functional material important for solving energy problem is a superconductor (SC). As well known, SCs exhibit zero-resistivity states at temperatures lower than their SC transition temperature ($T_c$). For most superconductors, SC states are emerged by formation of electron pairs called Cooper pairs, and the formation is achieved via electron-phonon interaction [21]. Therefore, the mechanisms of superconductivity mediated by phonon for most superconductors are classified as conventional type. Since 1986, high-$T_c$ SCs were discovered in Cu-based [22] and Fe-based [23] SCs, and their SC mechanisms have been believed to be unconventional, namely non-phonon-mediated ones. However, conventional mechanisms have been regarded as a promising way to achieve a high Tc due to the recent development of hydrogen-based SC materials under extremely high pressure [24,25]. In addition, lattice anharmonicity has been considered as a key factor for SC in hydrogen-based [26]. Therefore, the understanding of lattice anharmonicity in SCs is also an important issue. As well as in HEA-type TE materials, the effect of high entropy has recently been introduced in various SC [27–36]. In HEA-type SCs, the effect of disorder to electronic and phonon characteristics and the modification of anharmonicity have not been addressed yet. Therefore, the knowledge about the relationship between high-entropy states and anharmonicity would be useful for further material design of SCs with a higher $T_c$.

Motivated by those backgrounds reviewed above, we have studied anharmonicity of two-dimensional and three-dimensional systems with different configurational entropy of mixing from zero entropy to HEA states. To investigate the evolution of anharmonicity by the increase in configurational entropy, we used $\gamma_G$ in this study. As a conclusion, we propose that the anharmonicity in both two- and three-dimensional structures can be modified by the increase in configurational entropy. The suppression of anharmonicity in HEA states would be a common feature in HEA-type materials with different structural dimensionality.

### 1. 2. Grüneisen parameter ($\gamma_G$)

The $\gamma_G$ of inorganic materials is calculated using the following formula; $\gamma_G = \beta_V B V_{mol}/C_V$, where $\beta_V$, $B$, $V_{mol}$, and $C_V$ are volume thermal expansion coefficient, bulk modulus, molar volume, and specific heat, respectively. The parameters needed for the estimation of $\gamma_G$ are calculated as follows.

$$\beta_V = \frac{1}{V(300\ \mathrm{K})}\frac{dV}{dT} \tag{2}$$

$$B = \rho\left(v_L^2 - \frac{4}{3}v_s^2\right) \tag{3}$$

$$\theta_D = \left(\frac{h}{k_B}\right)\left[\frac{3n}{4\pi}\left(\frac{N_A\rho}{M}\right)\right]^{\frac{1}{3}} v_m \tag{4}$$

$$v_m = \left[\frac{1}{3}\left(\frac{2}{v_s^3} + \frac{1}{v_L^3}\right)\right]^{-\frac{1}{3}} \tag{5}$$

In the formulas, $dV/dT$, $\rho$, $v_L$, $v_S$, $v_m$, $\theta_D$, $h$, $k_B$, $n$, $N_A$, and $M$ denote temperature gradient of lattice volume, density of the material, longitudinal sound velocity, shear sound velocity, average sound velocity, Debye temperature, Plank's constant, Boltzmann's constant, number of atoms in the molecule (formula unit), Avogadro's constant, and the molecular weight (per formula unit). Although the absolute value of $\gamma_G$ of functional materials depends on crystal structure or constituent elements, at least, the value becomes a good scale for discussing the evolution of lattice anharmonicity by doping, pressurizing, or element substitution in similar compounds.

For example, in a metal telluride system $Pb_{1-x}Sn_xTe$, which is a TE material family [8] and a parent phase of SCs with topological electronic states [37,38], the solution of Sn and Pb results in an increase of $\gamma_G$: $\gamma_G$ = 1.5, 2.5, 2.8, and 2.1 for $x$ = 0, 0.25, 0.5, and 1 [39]. By the enhanced lattice anharmonicity in the alloy phase ($x$ = 0.25 and 0.5), $\kappa_{ph}$ is clearly suppressed, and the effect was explained by the changes in $\gamma_G$ [39]. In our previous study on $LaOBiS_{2-x}Se_x$, which is a TE system [15] and parent phases of layered SCs [40], we revealed that anharmonic lattice vibration is the origin of low $\kappa_{ph}$ using neutron inelastic scattering [12]. By partial Se substitution in $LaOBiS_{2-x}Se_x$, phonon softening is observed, and the $\kappa_{ph}$ decreases with decreasing the low-energy phonon energy. This trend was reproduced by the Se concentration dependence of $\gamma_G$ in $LaOBiS_{2-x}Se_x$ [41,42]. As described above, estimation of $\gamma_G$ is a good way to investigate the evolution of lattice anharmonicity in functional materials. Therefore, in this article, we investigate $\gamma_G$ for HEA-type compounds because the family has recently been drawing much attention as electronic materials including TE and SC materials. Here, we estimated $\gamma_G$ for a layered $BiS_2$-based system $REO_{0.7}F_{0.3}BiS_2$ with a HEA-type RE (rare-earth) site and MTe with a HEA-type M (metal) site.

## 2. Results

In our previous work on $RE(O,F)BiS_2$, we synthesized polycrystalline samples of $REO_{0.5}F_{0.5}BiS_2$ with different configurational entropy of mixing ($\Delta S_{mix}$) at the RE site [27], according to a compositional guideline established for alloy-based HEAs [43,44], where HEA composition is defined as one containing five or more elements with a composition range of 5–35 at% $\Delta S_{mix}$ is defined as $\Delta S_{mix} = -R\ \Sigma_i\ c_i \ln c_i$, where $c_i$ and $R$ are compositional ratio and the gas constant, respectively [44]. By changing the number of element and composition at the RE site, the samples with different $\Delta S_{mix}$ were systematically synthesized [47]. Notably, SC shielding fraction is improved via the suppression of in-plane Bi-S1 local disorder (local distortion) (see Fig. 1e for crystal structure image and the definition of the S1 site) by an increase in $\Delta S_{mix}$ in $REO_{0.5}F_{0.5}BiS_2$ [47]. Since all the examined samples had similar lattice constants, we concluded that the increase in $\Delta S_{mix}$ results in the modification of the local structure and SC properties in $REO_{0.5}F_{0.5}BiS_2$ [47]. Therefore, the estimation of $\gamma_G$ for RE(O,F)$BiS_2$ would be useful to understand the effect explained above.

**Table 1.** Sample information including nominal composition, $\Delta S_{mix}$ / $R$, relative density, sound velocity ($v_L$ and $v_S$), volume thermal expansion coefficient ($\beta_V$), Debye temperature ($\theta_D$), bulk modulus ($B$), and $\gamma_G$ .

| Sample no. | #1 | #2 | #3 | #4 |
|---|---|---|---|---|
| RE site | Pr | $La_{0.3}Pr_{0.4}Nd_{0.3}$ | $La_{0.2}Ce_{0.2}Pr_{0.25}Nd_{0.35}$ | $La_{0.2}Ce_{0.2}Pr_{0.2}Nd_{0.2}Sm_{0.2}$ |
| $\Delta S_{mix}$ / $R$ (RE) | 0 | 1.09 | 1.36 | 1.61 |
| Relative density | 99% | 97% | 97% | 99% |
| $V_L$ (m/s) | 3430 | 3400 | 3320 | 3260 |
| $V_S$ (m/s) | 1860 | 1730 | 1720 | 1850 |
| $\beta_V$ (1/K) | 0.0000369 | 0.0000356 | 0.0000365 | 0.0000388 |
| $\theta_D$ (K) | 221 | 207 | 205 | 219 |
| $B$ (GPa) | 47.9 | 54.4 | 50.9 | 41.8 |
| $\gamma_G$ | 0.94 | 1.02 | 0.98 | 0.86 |

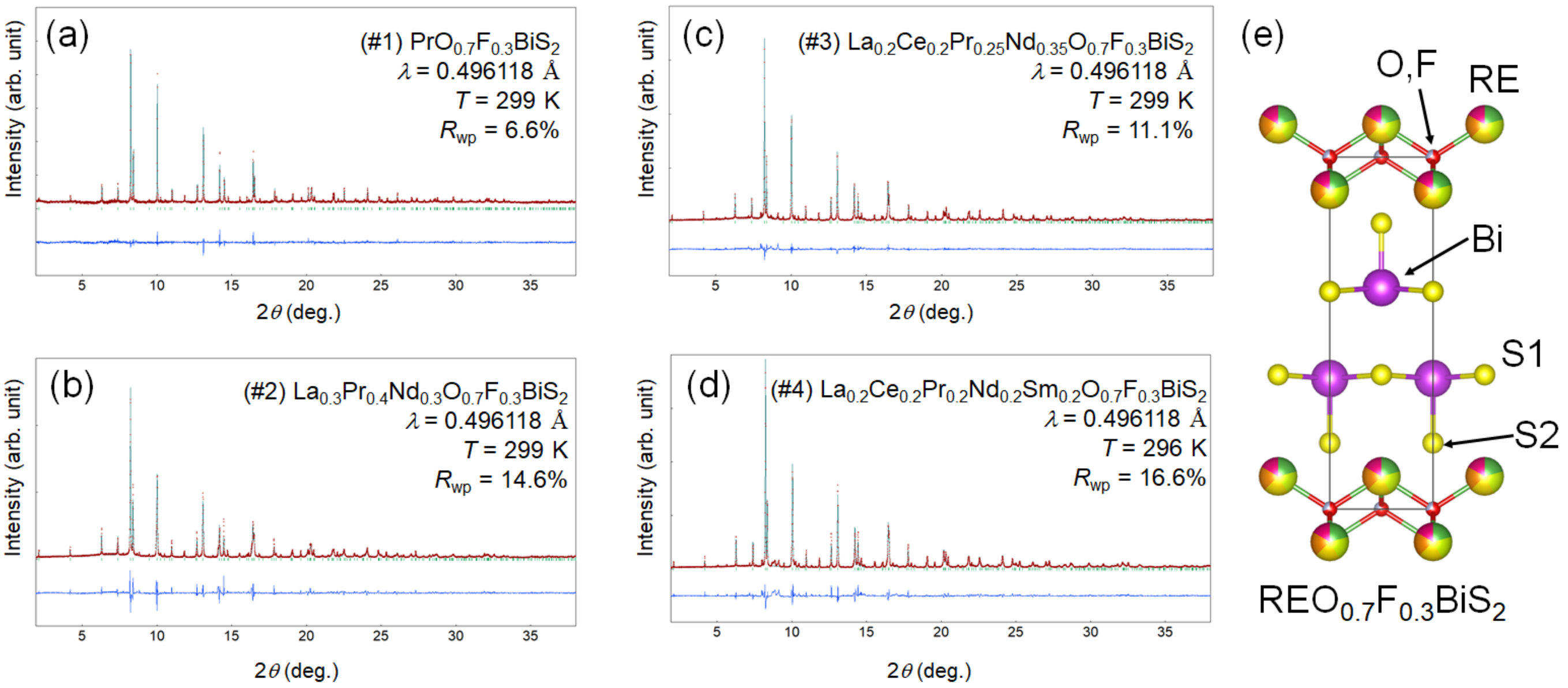

**Figure 1. SXRD patterns and Rietveld refinement results for (a) #1, (b) #2, (c) #3, and (d) #4. (e) Schematic image of crystal structure of $REO_{0.7}F_{0.3}BiS_2$.**

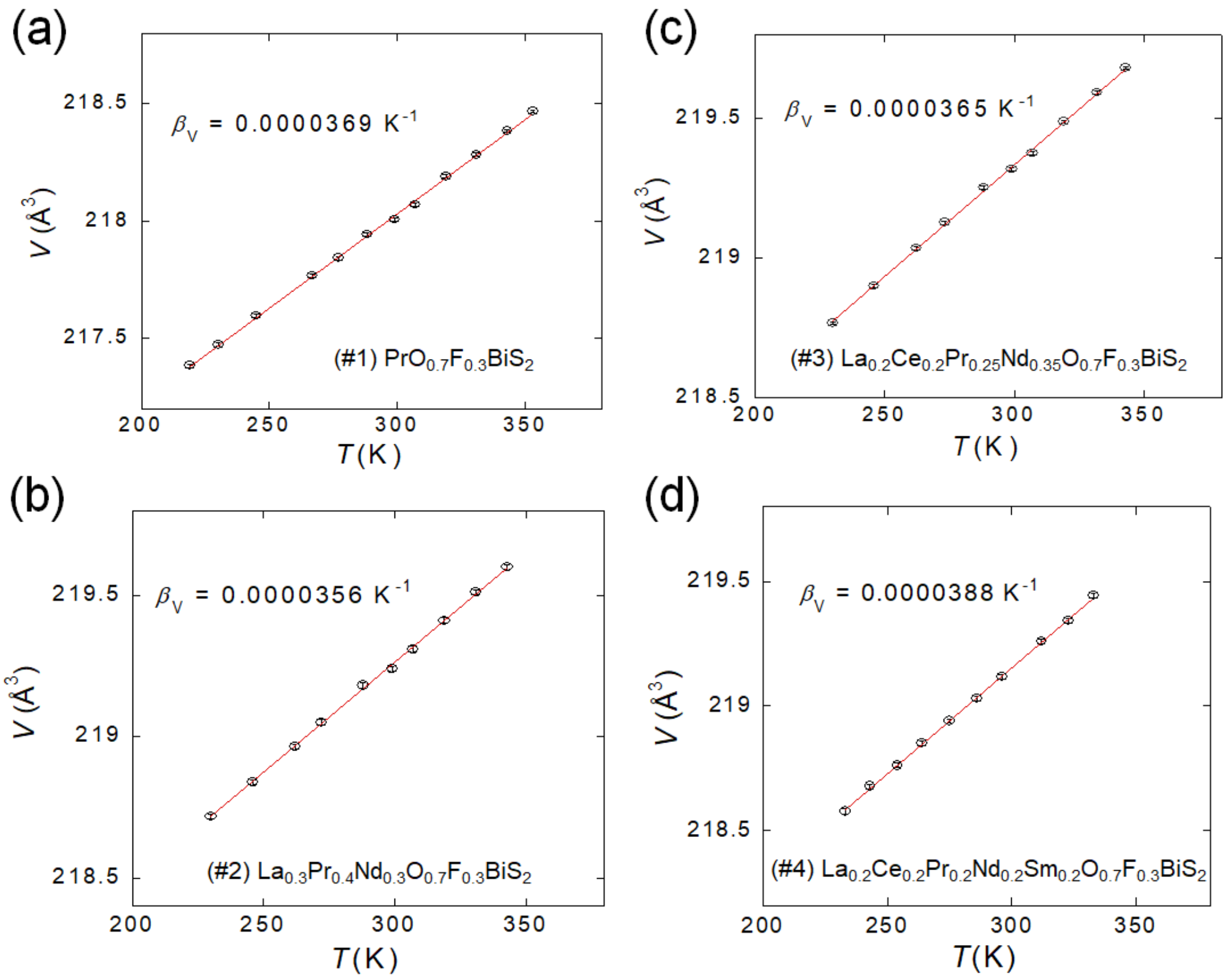

**Figure 2.** Temperature ($T$) dependence of lattice volume ($V$) for (a) #1, (b) #2, (c) #3, and (d) #4. The red lines show linear fitting results and the estimated $\beta_V$ is displayed.

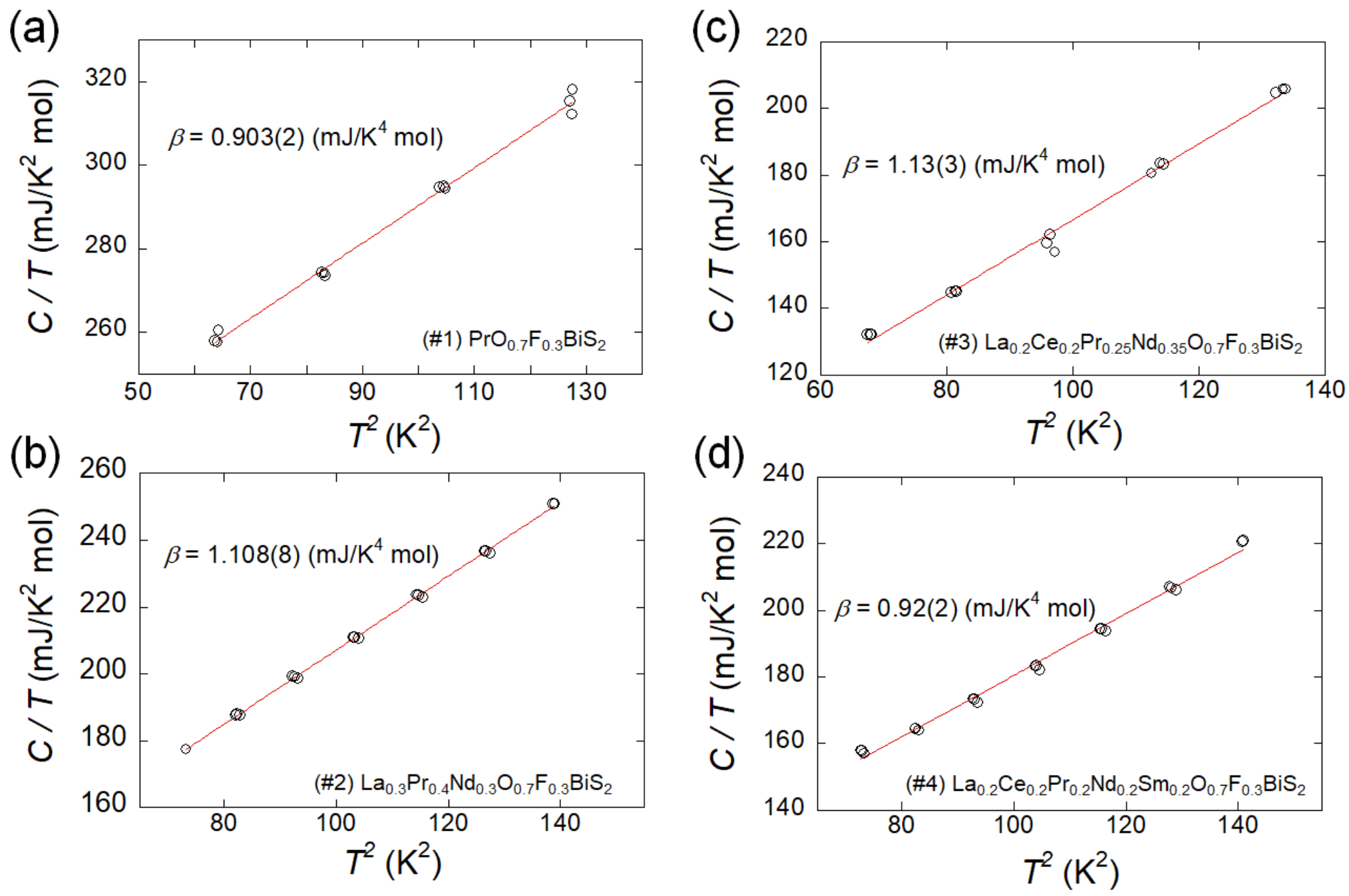

**Figure 3.** $T^2$ dependence of specific heat ($C/T$) for (a) #1, (b) #2, (c) #3, and (d) #4. The red lines show linear fitting results and the estimated phonon-contribution constant ($\beta$) is displayed.

Sample information for $REO_{0.7}F_{0.3}BiS_2$ (sample #1–#4) is listed in Table 1. By changing the RE-site elements, $\Delta S_{mix}$ has been systematically tuned, and sample #4 can be regarded as a HEA-type compound because of $\Delta S_{mix} > 1.5\ R$. The relative density and the values of $v_L$ were measured using high-pressure-annealed samples. We find that $v_L$ slightly decreases with increasing $\Delta S_{mix}$. To evaluate $\beta_V$, the temperature dependence of lattice volume was measured using synchrotron XRD (SXRD). Figure 1 shows the Rietveld refinement results on the SXRD patterns measured at temperatures near 300 K. Although small impurity peaks of $RE_2O_2S$ and/or $REF_3$ were found, the single-phase analysis was found to be sufficient for the estimation of lattice volume using the Rietveld refinement. The temperature dependence of the estimated lattice volume ($V$) is plotted in Fig. 2. The V linearly increases with increasing temperature. By linear fitting of the data and formula (2), $\beta_V$ was estimated and listed in Table 1.

Figure 3 shows the results of specific heat ($C$) for sample #1–#4. The data was analyzed based on a model for low-temperature region; $C = \gamma T + \beta T^3$, where $\gamma$ and $\beta$ are electronic and phonon-contribution specific heat constant, respectively. $\theta_D$ was calculated from $\beta$ using the following formula.

$$\beta = \frac{12\pi^4 N_A k_B}{5\theta_D^3} \tag{6}$$

$C_v$ was calculated using the Dulong–Petit law, which gives $C_v = 3R$. $V_{mol}$ was calculated using ideal density and molar mass estimated from the compositions. Those parameters needed for calculating $\gamma_G$ for $REO_{0.7}F_{0.3}BiS_2$ are listed in table 1. The estimated $\gamma_G$ is plotted in Fig. 4a. The $\gamma_G$ of #2 is larger than that of #1. The trend that $\gamma_G$ increases by little increasing $\Delta S_{mix.}$, is similar to the trend reported in $Pb_{1-x}Sn_xTe$ [39]. However, with further increasing $\Delta S_{mix.}$, $\gamma_G$ decreases in middle-to-high-entropy region. The results would suggest that the anharmonicity in $REO_{0.7}F_{0.3}BiS_2$ is enhanced in low-entropy region and is suppressed in middle- and HEA region. To explore commonality on this trend, we plotted the data presented in Ref. 39 in Fig. 4b and added $\gamma_G$ for HEA-type metal telluride ($AgInSnPbBiTe_5$) in the plot.

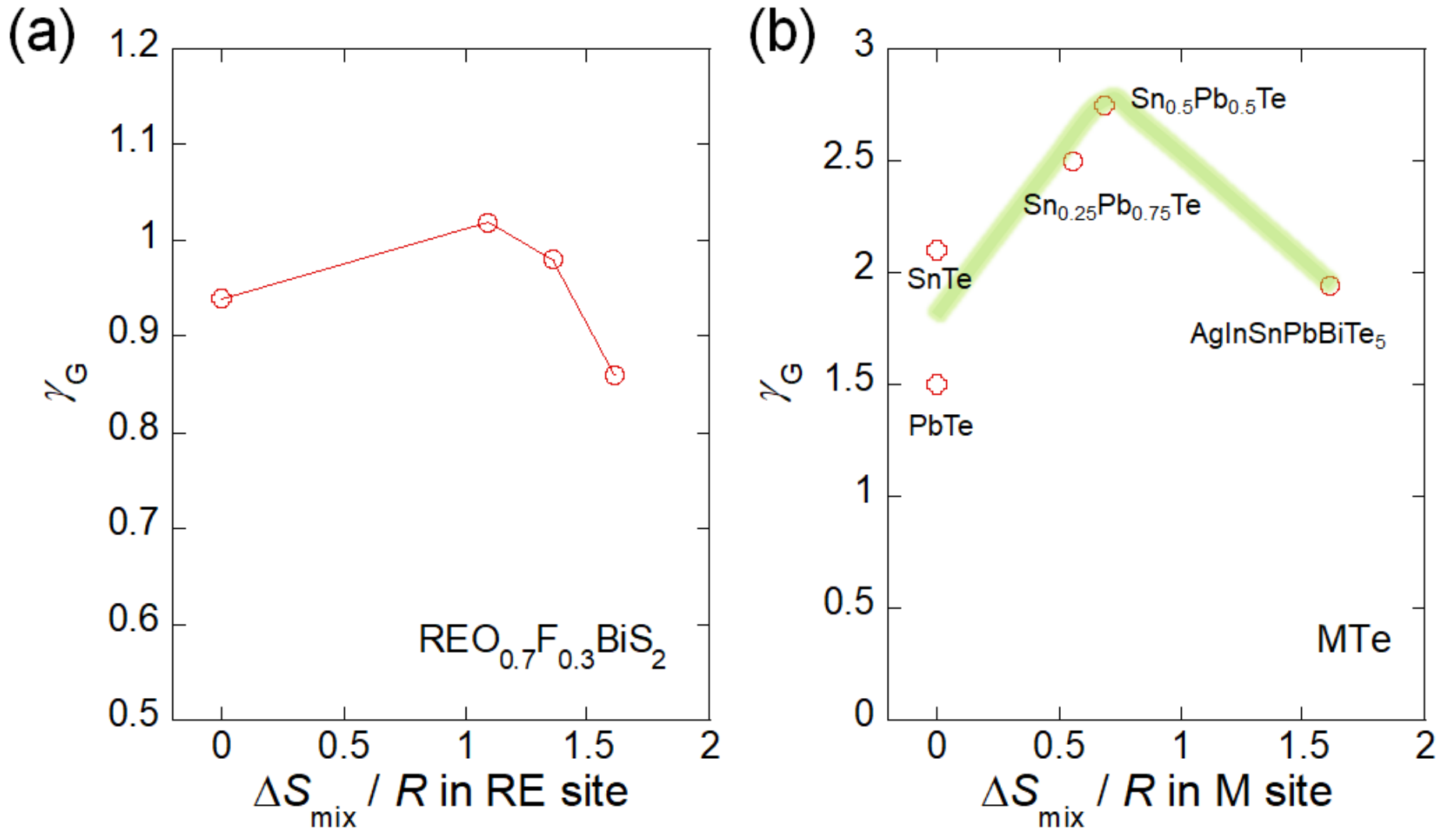

**Figure 4. Estimated $\gamma_G$ for (a) $REO_{0.7}F_{0.3}BiS_2$ and (b) MTe plotted as a function of $\Delta S_{mix}/R$. The data for $Pb_{1-x}Sn_xTe$ was taken from Ref. 39. Green line is eye guide.**

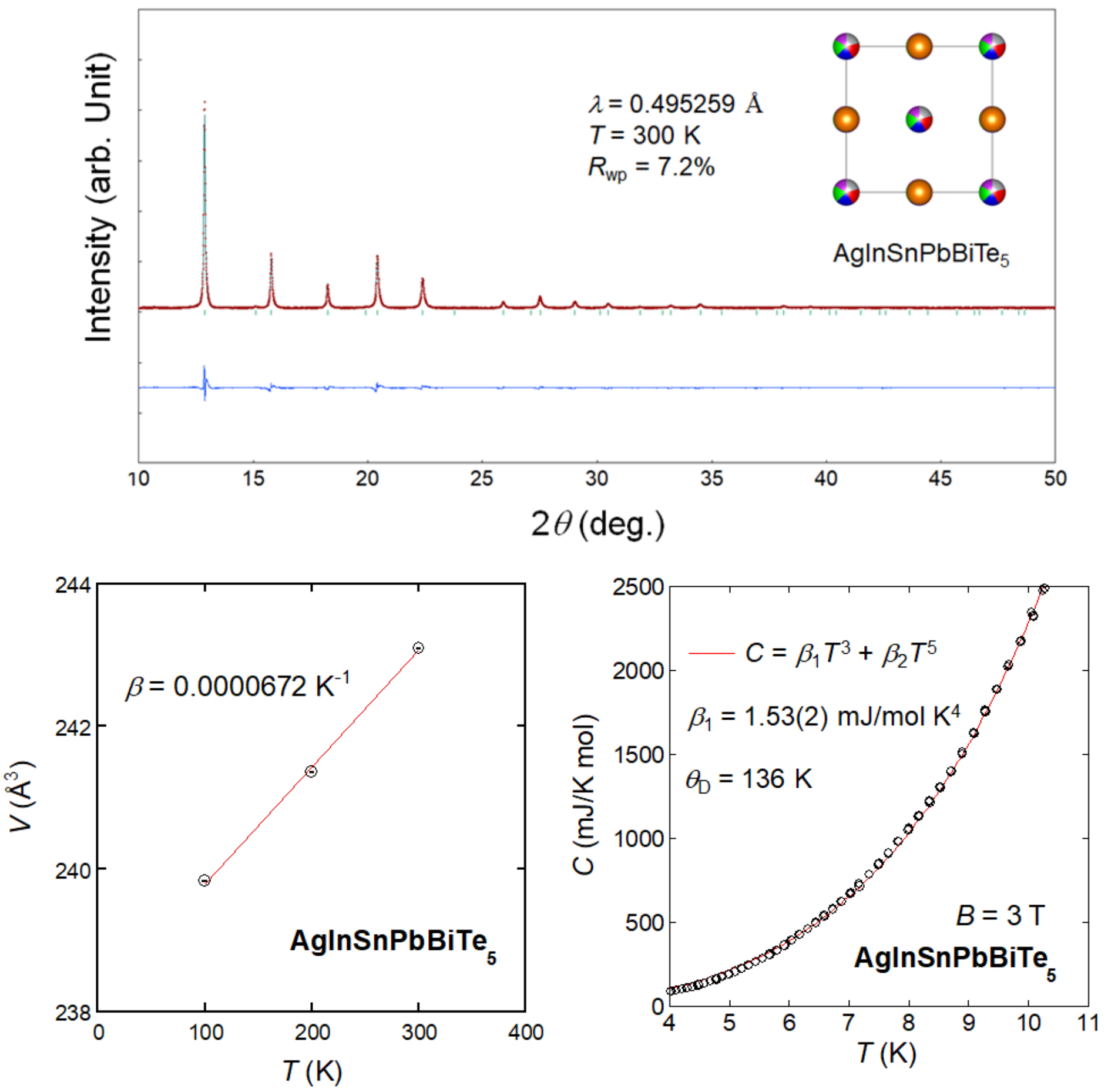

**Figure 5. (a) SXRD pattern and Rietveld refinement result for $AgInSnPbBiTe_5$. The inset shows schematic image of crystal structure. (b) Temperature dependence of $V$ for $AgInSnPbBiTe_5$. (c) Specific heat data for $AgInSnPbBiTe_5$ plotted as a function of $T$. The red line shows the fitting result.**

The temperature dependence of SXRD and low-temperature specific heat were measured on $AgInSnPbBiTe_5$. Figure 5a and 5b shows the results on Rietveld refinement on the SXRD pattern and the temperature evolution of $V$. The was estimated as $\beta_V$ = 0.0000672 $K^{-1}$. Figure 5c shows the low-temperature specific heat and the fitting result with a formula of $C = \beta_1 T^3 + \beta_2 T^5$, which was used in previous work on MTe [48]. Using $\beta_1$, $\theta_D$ is calculated as 136 K. The $v_L$ obtained for high-density (relative density of ~100%) was 2740 m/s. The calculated $B$ is 33.8 GPa, and the obtained $\gamma_G$ is 1.94. The trend on $\gamma_G$ for MTe (Fig. 4b) is quite interesting. As highlighted by the green line, $\gamma_G$ for MTe decreases in HEA region; the trend is common with Fig. 4a ($REO_{0.7}F_{0.3}BiS_2$). We consider that the trend would be caused by entropy tuning, and that is possibly a universal feature in HEA-type functional materials. In the next section, we briefly discuss the possible origin of the $\Delta S_{mix}$ dependence of $\gamma_G$.

## 3. Discussion

Entropy is a general concept to various physical quantity. To discuss the phenomena observed in the results of this work, we have to consider at least two different entropies, such as above-mentioned configurational entropy of mixing and vibrational entropy. In a field of glass transition, those configurational and vibrational entropies have been separately considered and analyzed, and the interplay has been discussed [47–49]. On the basis of such a concept established in metallic glasses, we discuss the origin of the results shown in Fig. 4. In low-to-middle-entropy region, the small disorder introduced by element substitution affect the average atomic positions, and displacements of the atoms and/or bonds should be weakly introduced. As revealed by extended X-ray absorption fine structure on $Pb_{1-x}Sn_xTe$ [39], bond anharmonicity is enhanced in the alloyed region, resulting in a large $\gamma_G$. Basically, structural disorder should be enhanced by an increase in $\Delta S_{mix}$ for both $REO_{0.7}F_{0.3}BiS_2$ and MTe systems. Therefore, in HEA region, the local structure (atomic positions and bonds) should approach glass-like states. In such a case, the total entropy in the system would be governed by the configurational entropy, which results in the suppression of the vibrational entropy. Although the results in this study are not exhibiting the direct evidence to the concept on the interplay between configurational and vibrational entropies, the concept would be useful to develop functional materials with lattice anharmonicity via entropy tuning. To establish this new concept, further investigation of $\gamma_G$ and lattice anharmonicity of functional materials are desired. Investigation on materials with harmonic lattice vibration is also needed.

## 4. Materials and Methods

The polycrystalline samples of $REO_{0.7}F_{0.3}BiS_2$ (see table 1 for nominal composition of the examined four samples with sample number #1–#4) were synthesized by a solid-state-reaction method. Powders of $La_2S_3$ (99.9%), $Ce_2S_3$ (99.9%), $Pr_2S_3$ (99.9%), $Nd_2S_3$ (99%), $Sm_2S_3$ (99.9%), $Bi_2O_3$ (99.999%), and $BiF_3$ (99.9%) and grains of Bi (99.999%) and S (99.99%) were used. The $Bi_2S_3$ powders were synthesized by reacting Bi and S in an evacuated quartz tube. The mixture of starting materials with a nominal composition was mixed by mortar and pestle, pelletized, and sealed into an evacuated quartz tube. The samples were heated at 700 °C for 20 h for #1 and at 750°C for 20 h for #2–#4. The obtained products were ground, pelletized, sealed into an evacuated quartz tube, and heated under the same heating conditions for homogenization. Since dense samples are needed for sound velocity measurement, the obtained $REO_{0.7}F_{0.3}BiS_2$ powders were annealed at 400°C for 15 minutes using a cubic-anvil-type high-pressure-synthesis system under 1.5 GPa. The obtained samples had a relative density higher than 97%. For the $AgInSnPbBiTe_5$ sample, precursor powder was synthesized by reacting Ag (99.9%) powder and In (99.99%), Sn (99.999%), Pb (99.9%), Bi, and Te (99.999%) grains with a nominal composition at 800 °C in an evacuated quartz tube. The obtained precursor was annealed at 500°C for 30 minutes under high pressure of 3 GPa.

The synchrotron XRD (SXRD) was performed at the beamline BL02B02, SPring-8 under proposals (Nos.: 2020A0068 and 2021B1175). The wavelength of the X-ray was 0.496118(1) Å for the experiments of $REO_{0.7}F_{0.3}BiS_2$ and 0.495259(1) Å for that for $AgInSnPbBiTe_5$. The SXRD experiments were performed with a sample rotator system; the diffraction data were collected using a high-resolution one-dimensional semiconductor detector (multiple MYTHEN system [50]) with a step size of $2\theta = 0.006°$. The temperatures of the samples were changed by $N_2$-gas temperature controller.

The crystal structure parameters were refined using the Rietveld method with the RIETAN-FP program [51]. The tetragonal *P*4/*nmm* (#129) model was used for the refinements for $REO_{0.7}F_{0.3}BiS_2$. For $AgInSnPbBiTe_5$, the NaCl-type (cubic *Fm*-3*m*; #225) model was used for the refinements. For structural parameters, we have reported in earlier works; for example, in Refs. 27, 28, and 40. The schematic images of crystal structure were drawn using VESTA software [52].

The temperature dependence of specific heat was measured on Physical Property Measurement System (PPMS, Quantum Design) by a relaxation method. Longitudinal sound velocity ($v_L$) of the sample was measured on dense samples using an ultrasonic thickness detector (Satotech-2000C). The $v_L$ was corrected using relative density of the polycrystalline sample [41], and the corrected $v_L$ was used for calculation of $\gamma_G$.

## 5. Conclusions

In the introduction part, characteristics of TE and SC materials and the importance of lattice anharmonicity in those materials were reviewed. Motivated by recent works on investigation of anharmonicity by $\gamma_G$, we investigated structural and physical properties of $REO_{0.7}F_{0.3}BiS_2$, which is a layered (two-dimensional) system, and NaCl-type (three-dimensional) metal telluride (MTe); for both systems, $\Delta S_{mix}$ was controlled by changing solution elements at the RE and M sites. By plotting the estimated $\Delta S_{mix}$, we found that an increase in $\Delta S_{mix}$ in low-to-middle-entropy region results in the enhancement of anharmonicity, but further increase in $\Delta S_{mix}$ in middle-to-high-entropy region clearly suppresses anharmonicity. Since the trend has been observed for both cases with two- and three-dimensional structures, we propose that the trend would be universal feature for functional materials in which configurational entropy of mixing is modified by alloying one or more sites. Further studies on $\gamma_G$ for various functional materials are desired to establish this concept, and that will open new pathway to material development by entropy tuning.

**Author Contributions:** Conceptualization, A.Y. and Y.M.; methodology, F.I.A., Y.N., K.T., R.M., C.M., and Y.M.; validation, A.Y., Y.G. and Y.M.; formal analysis, F.I.A., Y.N., A.Y., M.R.K., M.Y., Y.G., A.M., K.T., R.M., Y.T., C.M., and Y.M.; investigation, F.I.A., Y.N., A.Y., M.R.K., M.Y., Y.G., A.M., K.T., R.M., Y.T., C.M., and Y.M.; resources, Y.G., K.T., R.M., Y.T., C.M., and Y.M.; data curation, F.I.A., A.Y., M.R.K., K.T., R.M., C.M., and Y.M.; writing—original draft preparation, F.I.A. and Y.M.; writing—review and editing, F.I.A., Y.N., A.Y., M.R.K., M.Y., Y.G., A.M., K.T., R.M., Y.T., C.M., and Y.M.; visualization, F.I.A. and Y.M.; supervision, A.Y., Y.G., Y.T., and Y.M.; project administration, Y.G., K.T., C.M., and Y.M.; funding acquisition, Y. G. and Y.M. All authors have read and agreed to the published version of the manuscript.

**Funding:** This work was partially funded by Grant-in-Aid for Scientific Research (KAKENHI) (Nos. 18KK0076, 21K18834, 21H00151), JST-CREST (No. JPMJCR20Q4) and Tokyo Metropolitan Government Advanced Research (No. H31-1).

**Data Availability Statement:** The data reported in this article can be provided by corresponding author (Yoshikazu Mizuguchi) through reasonable requests.

**Acknowledgments:** The authors thank R. Kurita, M. Omprakash, and O. Miura for their supports in experiments and fruitful discussion.

**Conflicts of Interest:** The authors declare no conflict of interest.

## References

1. Snyder, G. J.; Toberer, E. S. Complex thermoelectric materials. *Nat. Mater.* 2008, 7, 105–114.
2. Jia, N.; Tan, X. Y.; Xu, J.; Yan, Q.; Kanatzidis, M. G. Achieving Enhanced Thermoelectric Performance in Multiphase Materials. *Acc. Mater. Res.* 2022, in printing (DOI: 10.1021/accountsmr.1c00228)
3. Franz, R.; Wiedemann, G. Ueber die Wärme-Leitungsfähigkeit der Metalle (German). *Annalen der Physik* 1853, 165, 497–531.
4. H. J. Goldsmid, proc. Phys. Soc. London 71, 633 (1958).
5. Terasaki, I.; Sasago, Y.; Uchinokura, K. Large thermoelectric power in $NaCo_2O_4$ single crystals. *Phys. Rev. B* 1997, 56, 12685–12687.
6. Chung, D. Y.; Hogan, T.; Brazis, P.; Lane, M. R.; Kannewurf, C.; Bastea, M.; Uher, C.; Kanatzidis, M. G. $CsBi_4Te_6$: A High-Performance Thermoelectric Material for Low-Temperature Applications. *Science* 2000, 287, 1024–1027.
7. Romanenko, A. I.; Chebanova, G. E.; Chen, T.; Su, W.; Wang, H. Review of the thermoelectric properties of layered oxides and chalcogenides. *J. Phys. D: Appl. Phys.* 2021, 55, 143001.

8. Pei, Y.; Shi, X.; LaLonde, A.; Wang, H.; Chen, H.; Snyder, G. J. Convergence of electronic bands for high performance bulk thermoelectrics. *Nature* 2011, 473, 66–69.
9. Takabatake, T.; Suekuni, K.; Nakayama, T.; Kaneshita, E. Phonon-glass electron-crystal thermoelectric clathrates: Experiments and theory. *Rev. Mod. Phys.* 2014, 86, 669–716.
10. Suekuni, K.; Lee, C. H.; Tanaka, H. I.; Nishibori, E.; Nakamura, A.; Kasai, H.; Mori, H.; Usui, H.; Ochi, M.; Hasegawa, T.; Nakamura, M.; Ohira-Kawamura, S.; Kikuchi, T.; Kaneko, K.; Nishiate, H.; Hashikuni, K.; Kosaka, Y.; Kuroki, K.; Takabatake, T. Retreat from Stress: Rattling in a Planar Coordination. *Adv. Mater.* 2018, 30, 1706230.
11. Li, C.; Hong, J.; May, A. F.; Bansal, D.; Chi, S.; Hong, T.; Ehlers, G.; Delaire, O. Orbitally driven giant phonon anharmonicity in SnSe. *Nat. Phys.* 2015, 11, 1063–1069.
12. Lee, C. H.; Nishida, A.; Hasegawa, T.; Nishiate, H.; Kunioka, H.; Ohira-Kawamura, S.; Nakamura, M.; Nakajima, K.; Mizuguchi, Y. Effect of rattling motion without cage structure on lattice thermal conductivity in $LaOBiS_{2-x}Se_x$. *Appl. Phys. Lett.* 2018, 112, 023903.
13. Zhao, L. D.; Lo, S. H.; Zhang, Y.; Sun, H.; Tan, G.; Uher, C.; Wolverton, C.; Dravid, V. P.; Kanatzidis, M. G. Ultralow thermal conductivity and high thermoelectric figure of merit in SnSe crystals. *Nature* 2014, 508, 373–377.
14. Suekuni, K.; Tsuruta, K.; Ariga, T.; Koyano, M. Thermoelectric Properties of Mineral Tetrahedrites $Cu_{10}Tr_2Sb_4S_{13}$ with Low Thermal Conductivity. *Appl. Phys. Express* 2012, 5, 051201.
15. Nishida, A.; Miura, O.; Lee, C. H.; Mizuguchi, Y. High thermoelectric performance and low thermal conductivity of densified LaOBiSSe. *Appl. Phys. Express* 2015, 8, 111801.
16. Jiang, B.; Yu, Y.; Cui, J.; Liu, X.; Xie, L.; Liao, J.; Zhang, Q.; Huang, Y.; Ning, S.; Jia, B.; Zhu, B.; Bai, S.; Chen, L.; Pennycook, S. J.; He, J. High-entropy-stabilized chalcogenides with high thermoelectric performance. *Science* 2021, 371, 6531.
17. Yamashita, A.; Goto, Y.; Miura, A.; Moriyoshi, C.; Kuroiwa, Y.; Mizuguchi, Y. n-Type thermoelectric metal chalcogenide (Ag,Pb,Bi)(S,Se,Te) designed by multi-site-type high-entropy alloying. *Mater. Res. Lett.* 2021, 9, 366.
18. Luo, Y.; Hao, S.; Cai, S.; Slade, T. J.; Luo, Z. Z.; Dravid, V. P.; Wolverton, C.; Yan, Q.; Kanatzidis, M. G. High Thermoelectric Performance in the New Cubic Semiconductor $AgSnSbSe_3$ by High-Entropy Engineering. *J. Am. Chem. Soc.* 2020, 142, 15187–15198.
19. Jiang, B. Yu, Y.; Chen, H.; Cui, J.; Liu, X.; Xie, L.; He, J. Entropy engineering promotes thermoelectric performance in p-type chalcogenides. *Nat. Commun.* 2021, 12, 3234.
20. Chen, K.; Zhang, R.; Bos, J. W. G.; Reece, M. J. Synthesis and thermoelectric properties of high-entropy half-Heusler $MFe_{1-x}Co_xSb$ (M = equimolar Ti, Zr, Hf, V, Nb, Ta). *J. Alloy. Compd.* 2022, 892, 162045.
21. Bardeen, J.; Cooper, L. N.; Schrieffer, J. R. Theory of Superconductivity. *Phys. Rev.* 1957, 108, 1175–1204.
22. Bednorz, J.G.; Müller, K.A. Possible high $T_c$ superconductivity in the Ba−La−Cu−O system. *Z. Physik B - Condensed Matter* 1986, 64, 189–193.
23. Kamihara, Y.; Watanabe, T.; Hirano, M.; Hosono, H. Iron-Based Layered Superconductor $La[O_{1-x}F_x]FeAs$ (x = 0.05−0.12) with $T_c$ = 26 K. J. Am. Chem. Soc. 2008, 130, 3296–3297.
24. Drozdov, A., Eremets, M., Troyan, I.; Ksenofontov, V.; Shylin, S. I. Conventional superconductivity at 203 kelvin at high pressures in the sulfur hydride system. *Nature* 2015, 525, 73–76.
25. Wang, H.; Li, X.; Gao, G.; Li, Y.; Ma, Y. Hydrogen-rich superconductors at high pressures. *WIREs Comput. Mol. Sci.* 2018, 8, e1330.
26. Setty, C.; Baggioli, M.; Zaccone, A. Anharmonic theory of superconductivity in the high-pressure materials. *Phys. Rev. B* 2021, 103, 094519.
27. Sogabe, R.; Goto, Y.; Mizuguchi, Y. Superconductivity in $REO_{0.5}F_{0.5}BiS_2$ with high-entropy-alloy-type blocking layers. *Appl. Phys. Express* 2018, 11, 053102.
28. Mizuguchi, Y. Superconductivity in High-Entropy-Alloy Telluride $AgInSnPbBiTe_5$. *J. Phys. Soc. Jpn.* 2019, 88, 124708.
29. Yamashita, A.; Jha, R.; Goto, Y.; Matsuda, T. D.; Aoki, Y.; Mizuguchi, Y. An efficient way of increasing the total entropy of mixing in high-entropy-alloy compounds: a case of NaCl-type $(Ag,In,Pb,Bi)Te_{1-x}Se_x$ (x = 0.0, 0.25, 0.5) superconductors. *Dalton Trans.* 2020, 49, 9118-9122.
30. Kasem, M. R.; Hoshi, K.; Jha, R.; Katsuno, M.; Yamashita, A.; Goto , Y.; Matsuda, T. D.; Aoki, Y.; Mizuguchi, Y. Superconducting properties of high-entropy-alloy tellurides M-Te (M: Ag, In, Cd, Sn, Sb, Pb, Bi) with a NaCl-type structure. *Appl. Phys. Express* 2020, 13, 033001.
31. Mizuguchi, M.; Kasem, M. R.; Matsuda, T. D. Superconductivity in $CuAl_2$-type $Co_{0.2}Ni_{0.1}Cu_{0.1}Rh_{0.3}Ir_{0.3}Zr_2$ with a high-entropy-alloy transition metal site. *Mater. Res. Lett.* 2021, 9, 141–147.
32. Kasem, M. R.; Yamashita, A.; Goto, Y.; Matsuda, T. D.; Mizuguchi, Y. Synthesis of high-entropy-alloy-type superconductors $(Fe,Co,Ni,Rh,Ir)Zr_2$ with tunable transition temperature. *J. Mater. Sci.* 2021, 56, 9499-9505.
33. Yamashita, A.; Matsuda, T. D.; Mizuguchi, Y. Synthesis of new high-entropy alloy-type $Nb_3$(Al, Sn, Ge, Ga, Si) superconductors. *J. Alloy Compd.* 2021, 868, 159233.
34. Shukunami, Y.; Yamashita, A.; Goto, Y.; Mizuguchi, Y. Synthesis of RE123 high-$T_c$ superconductors with a high-entropy-alloy-type RE site. *Phys. C* 2020, 572, 1353623.
35. Ying, T.; Yu, T.; Shiah, Y. S.; Li, C.; Li, J.; Qi, Y.; Hosono, H. High-Entropy van der Waals Materials Formed from Mixed Metal Dichalcogenides, Halides, and Phosphorus Trisulfides. *J. Am. Chem. Soc.* 2021, 143, 7042–7049.

36. Mizuguchi, Y.; Yamashita, A. Superconductivity in HEA-Type Compounds. In *Advances in High-Entropy Alloys - Materials Research, Exotic Properties and Applications*. Edited by Kitagawa, J. *IntechOpen* 2021, DOI: 10.5772/intechopen.96156.
37. Balakrishnan, G.; Bawden, L.; Cavendish, S.; Lees, M. R. Superconducting properties of the In-substituted topological crystalline insulator SnTe. *Phys. Rev. B* 2013, 87, 140507.
38. Mitobe, T.; Hoshi, K.; Kasem, M. R.; Kiyama, R.; Usui, H.; Yamashita, A.; Higashinaka, R.; Matsuda, T. D.; Aoki, Y.; Goto, Y.; Mizuguchi, Y. Superconductivity in In-doped $AgSnBiTe_3$ with possible band inversion. *Sci. Rep.* 2021, 11, 22885.
39. Knura, R.; Parashchuk, T.; Yoshiasa, A.; Wojciechowski, K. T. Origins of low lattice thermal conductivity of $Pb_{1-x}Sn_xTe$ alloys for thermoelectric applications. *Dalton Trans.* 2021, 50, 4323–4334.
40. Mizuguchi, Y. Material Development and Physical Properties of $BiS_2$-Based Layered Compounds. *J. Phys. Soc. Jpn.* 2019, 88, 041001.
41. Abbas, F. I.; Yamashita, A.; Hoshi, K.; Kiyama, R.; Kasem, M. R.; Goto, Y.; Mizuguchi, Y. Investigation of lattice anharmonicity in thermoelectric $LaOBiS_{2-x}Se_x$ through Gruneisen parameter. *Appl. Phys. Express* 2021, 14, 071002.
42. Abbas, F. I.; Yamashita, A.; Hoshi, K.; Kiyama, R.; Kasem, M. R.; Goto, Y.; Mizuguchi, Y. Corrigendum on Investigation of lattice anharmonicity in thermoelectric $LaOBiS_{2-x}Se_x$ through Gruneisen parameter. *Appl. Phys. Express* 2022, in printing.
43. Yeh, J. W.; Chen, S. K.; Lin, S. J.; Gan, J. Y.; Chin, T. S.; Shun, T. T.; Tsau, C. H.; Chang, S. Y. Nanostructured High-Entropy Alloys with Multiple Principal Elements: Novel Alloy Design Concepts and Outcomes. *Adv. Eng. Mater.* 2004, 6, 299–303.
44. Tsai, M. H.; Yeh, J. W. High-Entropy Alloys: A Critical Review. *Mater. Res. Lett.* 2014, 2, 107–123.
45. Sogabe, R.; Goto, Y.; Abe, T.; Moriyoshi, C.; Kuroiwa, Y.; Miura, A.; Tadanaga, K.; Mizuguchi, Y. Improvement of superconducting properties by high mixing entropy at blocking layers in $BiS_2$-based superconductor $REO_{0.5}F_{0.5}BiS_2$. *Solid State Commun.* 2019, 295, 43–49.
46. Katsuno, M.; Jha, R.; Hoshi, K.; Sogabe, R.; Goto, Y.; Mizuguchi, Y. High-Pressure Synthesis and Superconducting Properties of NaCl-Type $In_{1-x}Pb_xTe$ ($x$ = 0–0.8). *Condens. Matter* 2020, 5, 14.
47. Smith, H. L.; Li, C. W.; Hoff, A.; Garrett, G. R.; Kim, D. S.; Yang, F. C.; Lucas, M. S.; Swan-Wood, T.; Lin, J. Y. Y.; Stone, M. B.; Abernathy, D. L.; Demetriou, M. D.; Fultz, B. Separating the configurational and vibrational entropy contributions in metallic glasses. *Nat. Phys.* 2017, 13, 900 –905.
48. Gibbs, J. H.; DiMarzio, E. A. Nature of the glass transition and the glassy state. *J. Chem. Phys.* 1958, 28, 373–383.
49. Adam, G.; Gibbs, J. H. On the temperature dependence of cooperative relaxation properties in glass-forming liquids. *J. Chem. Phys.* 1965, 43, 139–146.
50. Kawaguchi, S.; Takemoto, M.; Osaka, K.; Nishibori, E.; Moriyoshi, C.; Kubota, Y.; Kuroiwa, Y.; Sugimoto, K. High-throughput powder diffraction measurement system consisting of multiple MYTHEN detectors at beamline BL02B2 of SPring-8. *Rev. Sci. Instrum.* 2017, 88, 085111.
51. Izumi, F.; Momma, K. Three-Dimensional Visualization in Powder Diffraction. *Solid State Phenom.* 2007, 130, 15–20.
52. Momma, K.; Izumi, F. VESTA: a three-dimensional visualization system for electronic and structural analysis. *J. Appl. Crystallogr.* 2008, 41, 653.